\documentstyle[12pt,a4]{article}
\def\ut#1{\lower1.2ex\hbox{$\mathchar"3218$}\mkern -14mu%
          \hbox to 2ex{\hss$#1$\hss}}
\def\utom{\lower1.2ex\hbox{$\mathchar"3218$}\mkern -14mu%
          \hbox to 2ex{\hss$\omega$\hss}}

\begin{document}

\title{Quantum states on supersymmmetric minisuperspace with
a cosmological constant}
\author{Robert Graham and Andr\'as Csord\'as\thanks{Permanent address:
Research Institute for Solid State Physics, P.O. Box 49, H1525 Budapest
Hungary}}
\date{
Fachbereich Physik, Universit\"at-Gesamthochschule Essen\\
45117 Essen\\ Germany}
\maketitle

\begin{abstract}
Spatially homogeneous models in quantum supergravity with a nonvanishing
cosmological constant are studied. A class of exact nontrivial solutions
of the supersymmetry and Lorentz constraints is obtained in terms of the
Chern-Simons action on the spatially homogeneous 3-manifold, both in
Ashketar variables where the solution is explicit up to reality conditions,
and, more concretely, in the tetrad-representation, where the solutions
are given as integral representations differing only by the contours of
integration.
In the limit of a vanishing cosmological constant earlier exact solutions
for Bianchi type IX models in the tetrad-representation are recovered and
additional asymmetric solutions are found.
\end{abstract}

In the present-day search for a consistent theory comprising both general
relativity and the quantum theory in appropriate limits one of the still
promissing lines of approach is the nonperturbative quantization of
gravity or supergravity. In order to explore many of the recent new ideas
developed in this approach \cite{1} spatially homogeneous 'minisuperspace'
models of the full theory have proved to be a very valuable tool, in
particular in gravity \cite{2}, less so, but for no fundamental
reasons, in supergravity \cite{3}-\cite{15}. In pure gravity spatially
homogeneous mini\-superspace models without or with anisotropy,
cosmological constant, or matter-coupling have been succesfully explored.
In supergravity, on the other hand, until recently only minisuperspace
models of the isotropic Friedmann type without or with cosmological
constant or scalar field coupling \cite{4} have been successfully
quantized. The reason is that the correct treatment of anisotropy in the
fermionic components of such supersymmetric models was only given very
recently in \cite{17,18}. There anisotropic spatially homogeneous
supersymmetric minisuperspace models restricted to the case without
cosmological constant or matter coupling were treated. An important
simplifying feature of this restriction is the fact that the fermion
number defined by the Rarita-Schwinger field then is a good quantum
number and each sector with a fixed fermion number may be treated
separately.

If a cosmological constant (or any form of matter) is allowed for \cite{4},
\cite{12}-\cite{14} all fermion sectors become coupled and have to be
considered simultaneously. Previous work on anisotropic supersymmetric
Bianchi class A models with a cosmological constant \cite{12}-\cite{14}
attempted to study the coupled fermion sectors but concluded that a
nontrivial quantum state solving all constraints of supergravity does not
exist, which is the same as stating that a consistent quantization of such
models is not possible. However, in the light of the recent results in
\cite{17,18} it is clear that the treatments of anisotropy in these
papers need revision and the problem must be considered open.

In the present paper we therefore study the quantization of supersymmetric
Bianchi class A models with a cosmological term. We shall, in fact, find
an analytic expression for a family of nontrivial quantum states with
components in all fermion sectors, satisfying all constraints. In the
limit of vanishing cosmological constant these states give back explicit
exact solutions obtained in \cite{18} for Bianchi type IX.

As shown in \cite{11} the study of minisuperspace models in supergravity
can be greatly simplified by the use of Ashketar-variables, which we shall
therefore employ to do most calculations. However, instead of applying
reality conditions to the Ashketar variables, we shall at the end
transform our results to the tetrad representation, which is more
easy to interpret, and which was used in all previous work [12-14].
The starting point is the Hamiltonian formulation of $N=1$-supergravity
with a cosmological term in Ashketar variables which was given by Jacobson
\cite{19}. For simplicity we shall adopt the notational conventions made
there.

In the tetrad representation the independent variables are the spatial
real components ${e_p}^{AA'}$ of the tetrad $(p\in1,2,3\,$;
$A,A'\in 1,2)$ and the Grassmannian components ${\psi_p}^A$ of the
Rarita-Schwinger field. The ${e_p}^{AA'}$ form the metric tensor
$h_{pq} = - {e_p}^{AA'}e_{qAA'}$ on the space-like homogeneity 3-surfaces
in the symmetric basis of 1-forms $\omega^p$, satisfying, in Bianchi
class-A models,
\[
d\omega^p = \frac{1}{2}m^{pq}\varepsilon_{qrs}\omega^r\wedge\omega^s
\]
where $\varepsilon_{qrs}$ is the spatial Levi-Civita tensor density.
We shall denote $h = det(h_{pq})$. The constant symmetric matrix $m^{pq}$
is defined by the particular Bianchi type chosen within the class A.
The volume of the homogeneous 3-surfaces (compactified, if necessary) is
denoted by $V=\int\omega^1\wedge\omega^2\wedge\omega^3$. Due to the choice
of a symmetric (non-coordinate) basis the ${e_p}^{AA'}$ and ${\psi_p}^A$
are independent of spatial coordinates.

In Ashketar variables, on the other hand \cite{19}, one first adds
a suitable complex term to the Lagrangean, which does not change the
equations of motion, and then uses the complexified spin-connection
$A_{pAB}=A_{p(AB)}$ and the tensor density $\tilde{\sigma}^{pAB}$
\begin{equation}
\sqrt{2}\tilde{\sigma}^{pAB}= -\varepsilon^{pqr}{e_q}^{AA'}{{e_r}^B}_{A'}
= \sqrt{2}\tilde{\sigma}^{p(AB)}
\label{eq:1}
\end{equation}
as a canonically conjugate pair of (complex) coordinates $A_{pAB}$ and
momenta $\tilde{\sigma}^{pAB}$. $\varepsilon^{pqr}$ numerically equals
$\varepsilon_{pqr}$. In view of the reality of ${e_p}^{AA'}$
the $\tilde{\sigma}^{pAB}$ must be Hermitian with respect to some
Hermitian matrix $n^{AA'}$ \cite{19}
\begin{equation}
\left(\tilde{\sigma}^{pAB}\right)^\dagger\equiv {n^A}_{A'}\,{n^B}_{B'}
\bar{\tilde{\sigma}}^{pA'B'} = {\tilde{\sigma}_p}^{AB}
\label{eq:2}
\end{equation}
where $n^{AA'}$ satisfies
\begin{equation}
n_{AA'}n^{AA'} = 2\,,\quad n_{AA'}{e_p}^{AA'} = 0
\label{eq:3}
\end{equation}
and is thereby determined (up to a sign) as a function of the $e_p^{AA'}$,
or the $\tilde{\sigma}^{pAB}$, if the latter is more convenient. In the
context
of supergravity this choice of variables has the additional advantage
that there are no second-class constraints and there is therefore no need
to introduce Dirac brackets. Thus the diffeomorphism-, Hamiltonian-,
Lorentz-, and supersymmetry constraints of the theory can be obtained
very directly \cite{19}, and are also easily reduced to our present,
spatially homogeneous case \cite{11,14}. In the following we shall only
need the Lorentz constraints and the supersymmetry constraints which imply
all others via the algebra of the symmetry generators. After canonical
quantization in the $\left(A_p^{AB}\,,\psi_p^A\right)$-representation
where ${\tilde{\sigma}^p}_{AB} = -\frac{1}{\sqrt{2}}
\partial/{\partial A_p}^{AB}$ they take the form
\begin{eqnarray}
J_{AB}\Psi &=& \frac{1}{\sqrt{2}}\left(-2{A_p}^C_{(A}
  \frac{\partial}{{\partial A_p}^{B)C}}
+ \psi_{p(A}\frac{\partial}{\partial {\psi_p}^{B)}}\right)\Psi = 0\\
\label{eq:4}
S^A\Psi &=& \frac{1}{\sqrt{2}}\left({A_p}^{AB}
\frac{\partial}{\partial{\psi_p}^B}
-2\sqrt{2}V^2mi\frac{\partial}{\partial A_{pAB}}\psi_{pB}\right)\Psi = 0\\
\label{eq:5}
 S^{\dagger A} &=& - \frac{1}{4}\varepsilon_{pqr}
 \frac{\partial}{\partial A_{pAB}}\frac{\partial}{\partial{{A_q}^B}_C}
  \left(Vm^{sr}\psi_{sC}
+ \varepsilon^{str}{A_{sC}}^D\psi_{tD}- 2\sqrt{2}im \frac{\partial}
{\partial{\psi_r}^C}\right) \Psi = 0
\label{eq:6}
\end{eqnarray}
Here, the scalar $2\sqrt{2}m$ denotes the cosmological constant in the
notation of \cite{19}. The operator ordering chosen in (6) is motivated
by the fact that polynomiality of this constraint in
$\frac{\partial}{\partial A_p^{AB}}$
was only achieved after supplying an additional (non-vanishing) factor
$h^{1/2}n_{AA'}$\cite{19}, which can only be supplied from the left.
The notation $S^{\dagger A}$ is not meant to imply that $S^{\dagger A}$
is the adjoint of $S^A$, because we leave open the problem of defining
a scalar product.

The physical wave-functions satisfying these constraints are to be
holomorphic in the complex variables $A_p^{AB}$. The transformation to the
tetrad-representation is achieved in two steps:
(i) The change from the $A_p^{AB}$-representation to the $e_p^{AA'}$-
representation is achieved by the generalized Fourier-transform
\begin{equation}
\Psi'({e_p}^{AA'}) = \int\left[\prod_p\prod_{(A\le B)}d{A_p}^{AB}\right]e
^{-{A_p}^{AB}{e_{qA}}^{A'}e_{rBA'}\varepsilon^{pqr}}\psi
 \left({A_p}^{AB}\right)
\label{eq:7}
\end{equation}
along a suitable 9-dimensional contour in the complex manifold spanned by
the $A_p^{AB}$ chosen in order to achieve convergence, and permitting
partial integration without boundary terms. Apart from this condition
the contour may still be chosen artbitrarily, and, indeed there are
different possible choices corresponding to different linearly
independent solutions \cite{20a}.

(ii) In order to undo the initial non-canonical complex transformation
of the Lagrangean an additional similarity tranformation has to be
performed \cite{20} which takes the form
\begin{equation}
\Psi({e_p}^{AA'}) = e^{-\phi({e_p}^{AA'})} \Psi'({e_p}^{AA'})
\label{eq:8}
\end{equation}
with
\begin{equation}
\phi = -\frac{V}{2}m^{pq}{e_p}^{AA'}e_{qAA'}=
 \frac{V}{2}m^{pq}h_{pq}\,.
\label{eq:9}
\end{equation}
Indeed, applying these transformations on the generators (4)-(6), and
contracting the generator (6) between the step (i) and (ii) with
$h^{-1/2}n_{AA'}$ from the left \cite{19} we obtain the Lorentz generator
$J_{AB}$ and the supersymmetry generators $S_A, \bar{S}_A'$ for the
class A Bianchi models in the tetrad representation \cite{17,18}, with a
cosmological term. The operator ordering obtained is that chosen in
\cite{18}. As long as $h\neq 0$, each step in these transformations can be
inverted, i.e. after specifying the integration contour in (\ref{eq:7})
the Ashketar representation and the tetrad-representation
are mathematically if not physically equivalent for our models (see
\cite{21a} for a similar discussion in the context of full gravity).

Let us now turn to a solution of the constraints (4)-(6). We start by
noting that (6) may be rewritten in terms of the function
$F(A_p^{AB}, \psi_p^A)$ defined by
\begin{equation}
F=-\frac{1}{4\sqrt{2}m}\left(Vm^{pq}{\psi_p}^A\psi_{qA}+
 \varepsilon^{pqr}{\psi_p}^A{A_{qA}}^B\psi_{rB}\right)
\label{eq:10}
\end{equation}
as
\begin{equation}
\varepsilon_{pqr}
\frac{\partial}{\partial A_{pAB}}\frac{\partial}{{{\partial A_q}^B}_C}
 \left(i \frac{\partial}{\partial{\psi_r}^C}+
\frac{\partial F}{\partial{\psi_r}^C}\right) \Psi = 0\,.
\label{eq:11}
\end{equation}
A special class of solutions (and we shall restrict ourselves to this
class in the following, even though, undoubtedly, more general solutions
do exist) is therefore
\begin{equation}
\Psi = \,\mbox{\rm const}\,\exp\left[i\left(F({A_p}^{AB},{\psi_p}^A)
+G({A_p}^{AB})\right)\right]
\label{eq:12}
\end{equation}
where $G$ is independent of the $\psi_p^A$ but otherwise arbitrary.
Choosing this function apropriately we can next satisfy the constraint (5).
This yields
\begin{equation}
G=\frac{i}{(4mV)^2}\left(
  Vm^{pq}{A_p}^{AB}A_{qAB}+\frac{2}{3}
   \varepsilon^{pqr}{A_p}^{AB}{A_{qB}}^C A_{rCA}\right)\,.
\label{eq:13}
\end{equation}
One can finally check that the Lorentz constraint (4) is fullfilled,
which is obvious because $F+G$ is a manifest Lorentz-scalar. The function
$G$ can be expressed by the Chern-Simons functional
integrated over the spatially homogeneous 3-manifolds. It is already known
that an exponential
of the Chern-Simons functional is a formal solution of canonically
quantized pure gravity with a cosmological term in Ashketar variables
\cite{20}. Furthermore, exponentials of supersymmetric extensions of the
Chern-Simons functional have also previously been obtained as
semi-classical WKB solutions of quantum supergravity \cite{21,22}. Here
we
find such a wave-function for our spatially homogeneous models as an exact
solution of all constraints. It seems a safe conjecture that even in full
supergravity with $\dot{a}$ cosmological constant an exact formal solution
of this form exists. We note, however, that the solution (\ref{eq:12}) is
not yet fully specified, as `reality conditions' for ${A_p}^{AB}$ still
need to be imposed. Instead of doing this we prefer to transform back
to the physically more transparent tetrad representation.

The transformation of the wave-function (\ref{eq:12}) with (\ref{eq:10}),
(\ref{eq:13}) from the Ashketar-representation to the metric
representation
(\ref{eq:7}), (\ref{eq:8}), (\ref{eq:9})
can be performed in the two steps
described above. The $A_p^{AB}$ integrals required in the first step
need a prior specification of the integration contour. Fortunately,
not all of these integrals need to be done, because only three of the nine
degrees of freedom of ${A_p}^{AB}$ are physical, while six correspond
to gauge freedoms (three from basis changes of the $\omega^p$, three
from Lorentz frame
rotations) which can be fixed by a choice of gauge and are not integrated
over in that gauge. However, even the remaining three integrals cannot
all be performed analytically. In the
limit of vanishing cosmological constant $m\to 0$ a stationary-phase
approximation becomes possible. As a preparation for performing this
approximation we expand the wave-function in the fermions
\begin{equation}
\Psi=\,\mbox{\rm const}\,\sum_{n=1}^3
 \frac{\left(iF({A_p}^{AB},{\psi_p}^A)\right)^n}{n!}
 e^{i G({A_p}^{AB})}\,.
\label{eq:14}
\end{equation}
One stationary phase point is at $A_p^{AB} = 0$ for all $p=1,2,3$ and
all $A,B\in 1,2$. The first solution is therefore defined by chosing
a suitable contour passing through this point. To discuss its limit
for $m\to0$ we need to keep only the dominant
fermion term for $m\to 0$ (which, because of the appearance of $m$ in
the denominator of (\ref{eq:10}), has 6 $\psi_p^A$ -factors and therefore
fermion number 6). Then we obtain from the stationary phase at
${A_p}^{AB} = 0$
\begin{equation}
\Psi({e_p}^{AA'})=\,\mbox{\rm const}\,
 \left(\prod_{p=1}^3\prod_{A=1}^2{\psi_p}^A\right)
  \exp\left(-\phi({e_p}^{AA'})\right)
\label{eq:15}
\end{equation}
where $\phi$ is defined in eq.(9). This is the well-known 'worm-hole
state' in the 6-fermion sector \cite{18}.
In the constant prefactor we have also absorbed the divergent factor
$m^{-3}$.

Other stationary phase points are at ${A_p}^{AB}\neq 0$ and further
solutions are obtained by chosing integration contours through any of
them \cite{20a}. To be specific we shall discuss this for the case of
Bianchi-type IX, but the other Bianchi types in class A can be treated
similarly, replacing the $S0(3)$ group by the respective Lie groups. For
Bianchi type IX we have
\[
\renewcommand{\arraystretch}{0.50}
  m^{pq}= \left(\begin{array}{ccc}1 & 0 & 0\\
                                  0 & 1 & 0\\
                                  0 & 0 & 1\end{array}
          \right)
\label{eq:16}
\]
Then there is a stationary phase at
\begin{equation}
{A_p}^{AB} = \frac{V}{2} {\tau_p}^{AB}
\label{eq:17}
\end{equation}
where the $\tau^{pAB}=m^{pq}{\tau_q}^{AB}$ are a basis of 2-dimensional
symmetric matrices satisfying ${\tau}^{pAB}=
\frac{1}{2} m^{pq}\varepsilon_{qrs}\tau^{rAC} {\tau^{sB}}_C$,
which determines
the ${\tau_p}^{AB}$ up to an orthogonal rotation of the basis. We choose
the explicit representation
\begin{equation}
\tau^{1AB} = {1\phantom{-}\,0\choose  0\phantom{-}\,1}^{AB}\,,\quad
\tau^{2AB} = {0\,\phantom{-}i\choose -i\,\phantom{-}0}^{AB}\,,\quad
\tau^{3AB} = {-i\,\phantom{-}0\choose 0 \, \phantom{-}i}^{AB}
\label{eq:18}
\end{equation}
and adopt the convention
\begin{equation}
\varepsilon^{AB}={\phantom{-}0\; 1\choose {-1}\,0}^{AB}\,,\quad
\varepsilon_{AB}={\phantom{-}0\;1 \choose {-1}\,0}_{AB}\,.
\label{eq:19}
\end{equation}

Let us now parametrize ${A_p}^{AB}$ by the two rotation matrices $P_{pq}$
(for coordinate rotations) and $Q_{pq}$(for Lorentz frame rotations), each
depending on three different Euler angles, which are gauge freedoms, and
by
the remaining three (still complex) elements $A_q$, which, after imposing
some reality conditions, are three physical degrees of freedom (one
combination of them playing the role of time):
\begin{equation}
{A_p}^{AB}=\sum_{q,r}P_{pq}A_qQ_{qr}\tau^{rAB}\,.
\label{eq:20}
\end{equation}
Here, and until further notice we drop the summation convention. As
expected $\Psi$ does not depend on $\ut{P}$, $\ut{Q}$. The canonical
momenta of the physical variables
$A_q = \frac{1}{2}\sum_{p,r}P_{pq}Q_{qr}{A_p}^{AB}{\tau^r}_{AB}$ are
$\sigma^q=-\frac{\partial}{\partial A_q}=\sqrt{2}\sum_{p,r}P_{pq}
Q_{qr}{\tilde{\sigma}^p}_{AB}\tau^{rAB}$.

Let us now fix the gauges by the choice $\ut{P}=1$, $\ut{Q}=1$, i.e.
by the condition $A_{pq}=0$, $p\ne q$. The remaining Fourier integral
over the $A_q$ then introduces the $\sigma^q$-representation for the
same gauge, and we define the variables $b_1$, $b_2$, $b_3$ by
$\sigma^1=b_2b_3$ and cyclic in 1, 2, 3. Considering the whole family of
gauges for fixed but arbitrary $C$-number rotation matrices $\ut{P}$,
$\ut{Q}$ we may define
\begin{equation}
{\tilde{\sigma}^p}_{AB} (\ut{P},\ut{Q})=
\frac{1}{2\sqrt{2}}\sum_{qr}P_{pq}\sigma^qQ_{qr}{\tau^r}_{AB}\,.
\label{eq:21}
\end{equation}
which coincides with the full operator ${\tilde{\sigma}^p}_{AB}$ up to
terms proportional to derivatives of the Euler angles contained in
$\ut{P}$ and $\ut{Q}$, i.e. to generators of the gauge group. The latter
are zero when acting on physical states, i.e. in Dirac's sense
\cite{25} we have the weak equality ${\tilde{\sigma}^p}_{AB}\approx
{\tilde{\sigma}^p}_{AB}(\ut{P},\ut{Q})$. For the
tetrad matrix ${e_p}^{AB}\equiv in^{BA'}{{e_p}^A}_{A'}$ and the
spatial metric this gives us the further weak equalities
\begin{eqnarray}
{e_p}^{AB}&\approx&\frac{1}{\sqrt{2}}\sum_{q,r}P_{pq}b_q
 Q_{qr}{\tau}^{rAB}\nonumber\\
h_{pq}    &\approx&\sum_rP_{pr}b_r^2P_{qr}
\label{eq:22}
\end{eqnarray}
which explains the significance of the $b_q$. Returning to the gauge
$\ut{P}=1$, $\ut{Q}=1$ (and also returning to the summation convention
where possible)
we may now write $\Psi'$ of eq.~(\ref{eq:7}) in the form
\begin{equation}
\Psi'(b_1,b_2,b_3)=\,\mbox{\rm const}\,\sum_n\frac{i^n}{n!}F^n
  \int_{(C)}\,dA_1dA_2dA_3(A_1A_2A_3)^2
    \exp\left[\sum_q A_q\sigma^q+iG(A_q)\right]\,.
\label{eq:23}
\end{equation}
The prefactor $(A_1A_2A_3)^2$ comes from the Jacobian of the gauge fixing.
We have introduced the abbreviations
\begin{eqnarray}
F&=&-\frac{1}{4\sqrt{2}m}\Bigg(Vm^{pq}{\psi_p}^A\psi_{qA}-\sum_q
  \varepsilon^{pqr}{\psi_p}^A{\tau^q}_{AB}{\psi_r}^B
   \frac{\partial}{\partial\sigma^q}\Bigg)\nonumber\\
G(A_q)&=&\frac{1}{2m^2V^2}\left(\frac{V}{4}(A_1^2+A_2^2+A_3^2)-
          A_1A_2A_3\right)\,.
\label{eq:24}
\end{eqnarray}
The 3-dimensional contour $C$ in the complex $A_1$, $A_2$, $A_3$ manifold
now remains to be chosen. Points of stationary phase, for $m\to 0$,
satisfy the equations
\begin{equation}
 2A_1A_2=VA_3\quad\mbox{\rm and cyclic}
\label{eq:25}
\end{equation}
with solutions
\begin{eqnarray}
 (\phantom{ii}{\rm i})\;A_1&=&A_2=A_3=0\;,\;({\rm ii})
                              A_1=A_2=A_3=\frac{V}{2};\nonumber\\
 {\rm (iii)}\;A_1&=&A_2=-\frac{V}{2}\;,\; A_3=\frac{V}{2};\nonumber\\
(\phantom{i}{\rm iv})\; A_2&=&A_3=-\frac{V}{2}\;,\;A_2=
                                            \frac{V}{2};   \nonumber\\
(\phantom{ii}{\rm v})\; A_3&=&A_1=-\frac{V}{2}\;,\;A_2=
                                    \frac{V}{2};
\label{eq:26}
\end{eqnarray}
Choosing the contour $C$ to run through any of these stationary points
we generate a family of five linearly independent solutions \cite{20a}.
We discuss them briefly in the asymptotic limit $m\to0$ where we now also
use eq.~(\ref{eq:8}). The first state, arising from the choice (i),
has already been discussed above. In the stationary point the Jacobian
in (\ref{eq:21}) is singular and it is better to use (\ref{eq:7}) in
this case.
The second state with choice (ii) yields
\begin{equation}
\psi(b_1,b_2,b_3)\simeq\;\mbox{\rm const}\;
  e^{-\frac{V}{2}(b_1^2,+b_2^2+b_3^2)+V(b_1b_2+b_2b_3+b_3b_1)}
   F_0^2
\label{eq:27}
\end{equation}
with
\begin{equation}
F_0=-\frac{V}{4\sqrt{2}m}\left(m^{pq}{\psi_p}^A\psi_{qA}
    -\frac{1}{2}\sum_q\varepsilon^{pqr}{\psi_p}^A{\tau^q}_{AB}
     {\psi_r}^B\right)\,.
\label{eq:28}
\end{equation}
After some algebra it turns out that the term $F_0^3$ vanishes and
does not appear in eq.~(\ref{eq:27}), even though, if nonzero, it would
be the dominant term. The state (\ref{eq:27}) is the Hartle-Hawking state
in the 4-fermion sector, first found in \cite{19} by a completely
different approach.

The states arising from the choices (iii) to (v) are asymmetric but
related by cyclic permutations of the coordinate directions 1, 2, 3.
It is therefore sufficient to consider choice (iii) only, where we obtain
asymptotically
\begin{equation}
\psi(b_1,b_2,b_3)\simeq\;\mbox{\rm const}\;
  e^{-\frac{V}{2}(b_1^2,+b_2^2+b_3^2)+V(b_1b_2-b_2b_3-b_3b_1)}
   F_1^2
\label{eq:29}
\end{equation}
with
\begin{equation}
F_1=-\frac{V}{4\sqrt{2}m}\left(m^{pq}{\psi_p}^A\psi_{qA}
  -\frac{1}{2}\sum_q\varepsilon^{pqr}{\psi_p}^A
  \Bigg({\displaystyle-1}_{{\displaystyle-1}_{\displaystyle+1}}\Bigg)_{qs}
      {\tau^s}_{AB}{\psi_r}^B\right)\,.
\label{eq:30}
\end{equation}
(\ref{eq:29}) is also a state in the 4-fermion sector. This state could
have, but in fact has not been discussed before.

In summary, we have obtained a special family of solutions of the quantized
constraints of the supersymmetric Bianchi type models in class A with
a nonvanishing cosmological constant. In Ashketar variables these
solutions are all given by the exponential of a supersymmetric extension
of the complex Chern-Simons functional, restricted to the homogeneous
spatial 3-manifold under study. The different solutions of the family
arise by transforming back to the tetrad variables after fixing
a diffeomorphism and Lorentz gauge and using different
integration contours C. In stationary phase approximation for $m\to0$
we recover our earlier results, obtained for $m\to0$, and find three
additional asymmetric solutions in the 4-fermion sector. It is
obvious that in addition to the special family more general solutions
of the quantized constraints also
exist. Our result is in contrast to earlier work
\cite{12}-\cite{14} which, on the basis of an overly restrictive ansatz
for the wave-function, concluded that for these anisotropic models the
quantized constraints have only the trivial solution.

A lot remains to be done, even within the restrictions of these spatially
homogeneous anisotropic models: What is the significance of the appearance
of the Chern-Simons functional? Can one find more general
{\it analytical} solutions? Is it possible to place a scalar product
and a Hilbert space structure on the space of solutions? Can one find
solutions for anisotropic homogeneous models including matter? To these
and related questions we hope to return in future work.

\vspace{0.5cm}

\noindent
{\Large\bf Acknowledgements}

This work has been supported by the Deutsche Forschungsgemeinschaft
through the Sonderforschungsbereich 237 ``Unordnung und gro{\ss}e
Fluktuationen''. One of us (A.~Csord\'as) would like to acknowledge
additional support by The Hungarian National Scientific Research
Foundation under Grant number F4472.

\end{document}